# Optimizing Multi-Gateway LoRaWAN via Cloud-Edge Collaboration and Knowledge Distillation

Hong Yang
*College of Computer and Data Science*
*Fuzhou University*
Fuzhou, China
hongyang0066@gmail.com

*Abstract*—For large-scale multi-gateway LoRaWAN networks, this study proposes a cloud-edge collaborative resource allocation and decision-making method based on edge intelligence, HEAT-LDL (HEAT-Local Distill Lyapunov), which realizes collaborative decision-making between gateways and terminal nodes. HEAT-LDL combines the Actor-Critic architecture and the Lyapunov optimization method to achieve intelligent downlink control and gateway load balancing. When the signal quality is good, the network server uses the HEAT algorithm to schedule the terminal nodes. To improve the efficiency of autonomous decision-making of terminal nodes, HEAT-LDL performs cloud-edge knowledge distillation on the HEAT teacher model on the terminal node side. When the downlink decision instruction is lost, the terminal node uses the student model and the edge decider based on prior knowledge and local history to make collaborative autonomous decisions. Simulation experiments show that compared with the optimal results of all compared algorithms, HEAT-LDL improves the packet success rate and energy efficiency by 20.5% and 88.1%, respectively.

# Introduction

As the scale of the network continues to increase, the downlink capacity of multiple gateways is still limited, which will lead to downlink starvation problems. Terminal nodes will use poor strategies to send uplink messages for a long time due to the inability to receive downlink control messages from the gateway, which leads to a decrease in network performance. In addition, multiple gateways can avoid downlink overload of a single gateway through load balancing. When a device can access multiple gateways at the same time, how to select the optimal gateway for downlink control message transmission to maximize network performance under the condition of limited downlink links. In response to the above problems, this section proposes a cloud-edge collaborative resource allocation and decision-making method based on edge intelligence, and verifies the effectiveness of this method through a series of experimental evaluations.

# Cloud-Edge Collaborative Resource Allocation and Decision-making Method Based on Edge Intelligence

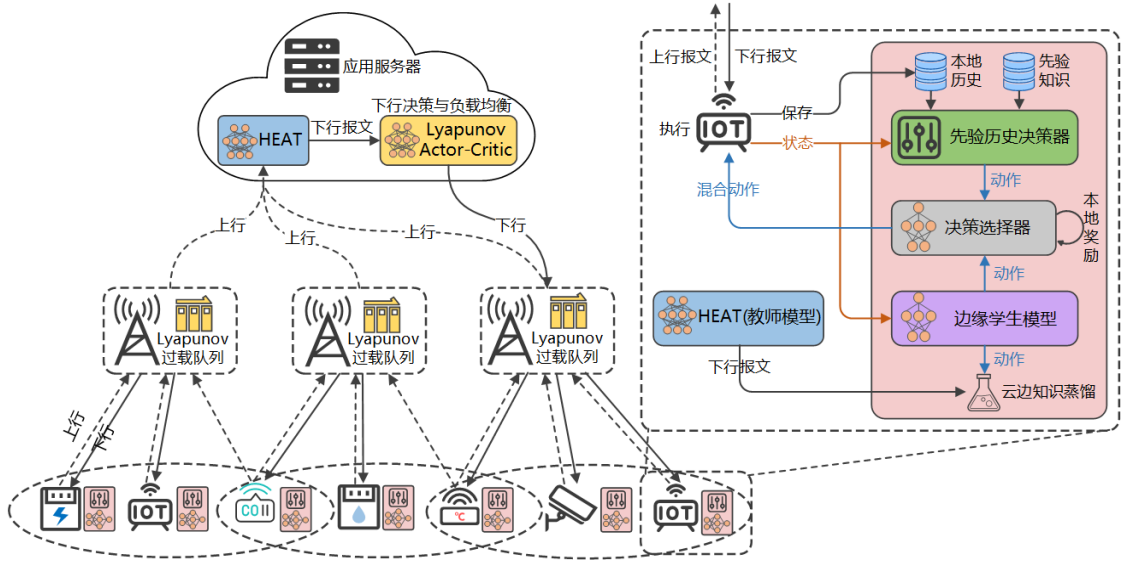

Fig.1 Overall framework of cloud-edge collaborative resource allocation and decision-making method based on edge intelligence

　　The proposed cloud-edge collaborative resource allocation and decision-making method based on edge intelligence mainly includes the following four modules: the HEAT algorithm proposed in Chapter 4, the edge decision module based on prior knowledge and local history, the online cloud-edge knowledge distillation module, and the intelligent downlink control and load balancing module based on Lyapunov optimization. Therefore, the proposed method is referred to as HEAT-LDL (HEAT-Local Distill Lyapunov). The overall framework diagram of HEAT-LDL is shown in Fig.1. From the overall framework diagram, the HEAT-LDL method is deployed in the three layers of cloud, gateway and terminal node, and forms a top-down and bottom-up two-way feedback. Among them, the HEAT algorithm is responsible for generating the original downlink control message. The Lyapunov Actor-Critic module deployed in the cloud and the Lyapunov overload queue of the gateway jointly realize the output intelligent downlink control and load balancing. The edge decision module and the knowledge distillation module are deployed on the terminal node. In the scenario where there is a lack of downlink control messages, the two make collaborative decisions to ensure the robustness and efficiency of the network.

　　Through the organic combination of these three modules, the HEAT-LDL method can not only achieve global optimization of multi-gateway loads in the cloud, but also make full use of the local prior knowledge of the terminal nodes and the distilled lightweight model to realize cloud-edge collaborative resource allocation and decision-making in multi-gateway large-scale LoRaWAN networks, while improving the overall network throughput and reducing latency, maintaining the scalability and robustness of the system.

## 2.1 Edge Decision Making Based on Prior Knowledge and Local History

The parameter decision problem of the terminal node $s^i$ at $t$ can be modeled as a joint optimization problem of a series of discrete random variables $\mathbf{P}_1$, as follows:

$$\mathbf{P}_1: \underset{\{u,p,d,w\}}{\mathrm{argmax}}\ \mathbb{P}(U=u, P=p, D=d, W=w) \tag{1}$$

$$\begin{aligned}
\text{s.t.}\ &(a): \{U=u\,|\,u \in \mathcal{A}_{USF}\},\\
&(b): \{P=p\,|\,p \in \mathcal{A}_P\},\\
&(c): \{D=d\,|\,d \in \mathcal{A}_{DSF}\},\\
&(d): \{W=w\,|\,w \in \mathcal{A}_w\}.
\end{aligned}$$

Among them, $U$, $P$, $D$ and $W$ are all random variables, corresponding to the uplink spreading factor, transmit power, downlink spreading factor and receiving window size of the terminal node respectively. $\mathcal{A} = \{\mathcal{A}_{USF}, \mathcal{A}_P, \mathcal{A}_{DSF}, \mathcal{A}_w\}$, where $\mathcal{A}_{USF}$, $\mathcal{A}_P$, $\mathcal{A}_{DSF}$ and $\mathcal{A}_w$ represent the action subspaces of each parameter in the total action space $\mathcal{A}$. Constraints (a) to (d) correspond to the range of values of the random variable $U, P, D, W$ that must meet the range specified in the LoRaWAN protocol. $\mathbb{P}$ is the probability that the current terminal node takes $\{u, p, d, w\}$ as the decision action. This value reflects the performance of the action combination $\{u, p, d, w\}$. The larger the $\mathbb{P}(u, p, d, w)$, the higher the PDR and EER of the action combination.

However, due to the unequal physical resources of the terminal node and the gateway and the constraints of the LoRaWAN protocol, the self-decision at the terminal node will be subject to additional restrictions. For the uplink, since the LoRa gateway chip equipped with the LoRaWAN gateway can detect and demodulate the signals corresponding to all spreading factors in parallel on 8 uplink channels, the terminal node does not need to negotiate with the gateway, and the terminal node can use any spreading factor to send uplink messages. For the downlink, the resource-constrained terminal node usually has only one channel and can only detect a specific spreading factor at the same time. Therefore, if the terminal node arbitrarily modifies its own downlink spreading factor $D$, this will cause the terminal node to be unable to demodulate the downlink message of the gateway. According to the LoRaWAN protocol, the downlink message of the gateway must be sent within the receiving window of the terminal node. Similarly, the terminal node arbitrarily modifies its own receiving window size $W$ may cause the terminal node to close the receiving window prematurely and miss the downlink message of the gateway. Due to the above additional constraints, the original optimization problem $\mathbf{P}_1$ can be transformed into a new optimization problem $\mathbf{P}_2$:

$$\mathbf{P}_2: \underset{\{u,p\}}{\mathrm{argmax}}\ \mathbb{P}(U=u, P=p\,|\,D=d, W=w) \tag{2}$$

$$\begin{aligned}
\text{s.t.}\ &(a): \{U=u\,|\,u \in \mathcal{A}_{USF}\},\\
&(b): \{P=p\,|\,p \in \mathcal{A}_P\}.
\end{aligned}$$

Although $\mathbf{P}_2$ is an integer nonlinear programming problem, the feasible solution space of $\mathbf{P}_2$ is small, only $|\mathcal{A}_{USF}| \cdot |\mathcal{A}_P| = 48$, so the optimal solution can be directly obtained by enumeration. So far, there is only one problem left to be solved, that is, how to obtain the $\mathbb{P}(u,p\,|\,d,w)$ corresponding to the action combination $\{u,p\}$.

To this end, the local historical data of the terminal node can be used to approximate $\mathbb{P}(u,p\,|\,d,w)$ under the time slice $T$ as follows:

$$\hat{\mathbb{P}}(TX_T\,|\,RX_T) = \begin{cases} 1 & , T=0; \\ N_s(TX_T, RX_T) + \rho \mathbb{P}(TX_{T-1}\,|\,RX_{T-1}) & , T>0. \end{cases} \quad (3)$$

$\hat{\mathbb{P}}$ is an approximation of $\mathbb{P}$, where $TX_T = \{u_T, p_T\}$, $RX_T = \{d_T, w_T\}$. $N_s(TX_T, RX_T)$ is the number of uplink messages successfully demodulated in time slice $T$ using parameter combination $\{TX_T, RX_T\}$. $\rho \in (0,1)$ is used to weigh short-term and long-term rewards. A smaller $\rho$ value makes the terminal node pay more attention to the action combination with a high current success rate, while a larger $\rho$ value makes the terminal node pay more attention to the action combination with a high long-term success rate.

However, due to the sparse traffic of terminal nodes in LoRaWAN networks, there are still some problems in using $\hat{\mathbb{P}}$ as an approximation of $\mathbb{P}$. LoRaWAN usually operates in unlicensed frequency bands, which are often subject to strict duty cycle restrictions. This restriction requires that terminal nodes can only have extremely short transmission periods per unit time, forcing a longer data transmission interval. In addition, since most terminal nodes are resource-constrained devices, frequent wireless transmissions will significantly increase energy consumption, so the transmission frequency of terminal nodes is usually low. This means that it takes a long time for terminal nodes to collect more historical data for $\hat{\mathbb{P}}$ to approximate $\mathbb{P}$. In the early stages of operation, the strategies derived from $\hat{\mathbb{P}}$ are close to uniform distribution, which leads to inefficient decision-making.

Introducing prior knowledge into the model is a common method to solve the problem of insufficient data. Prior knowledge can be combined with data-driven learning to achieve better performance without relying entirely on large-scale data, and is suitable for complex scenarios in actual engineering. Therefore, more prior knowledge can be introduced in $\hat{\mathbb{P}}$ to improve decision-making efficiency. Since the joint impact of $TX$ and $RX$ on network performance is more complicated to analyze, it is difficult to directly introduce prior information about $\mathbb{P}(TX\,|\,RX)$ and $\mathbb{P}(RX\,|\,TX)$. However, the independent impact of $TX$ and $RX$ on network performance is relatively simple to analyze, so the prior information about $TX$ and $RX$ is introduced in $\hat{\mathbb{P}}$ respectively. First, transform $\mathbb{P}(TX\,|\,RX)$ as follows:

$$\begin{aligned} \mathbb{P}(TX\,|\,RX) &= \mathbb{P}(RX\,|\,TX) \frac{\mathbb{P}(TX)}{\mathbb{P}(RX)} \\ &\approx \mathbb{P}(RX\,|\,TX) \frac{\mathbb{P}(TX) + \varrho_{\mathrm{tx}} \cdot \mathcal{U}_{\mathrm{tx}}(TX)}{\mathbb{P}(RX) + \varrho_{\mathrm{rx}} \cdot \mathcal{U}_{\mathrm{rx}}(RX)}. \end{aligned} \quad (4)$$

Among them, $\mathcal{U}_{\mathrm{tx}}(TX)$ and $\mathcal{U}_{\mathrm{rx}}(RX)$ are the prior information of $TX$ and $RX$, respectively, and $\varrho_{\mathrm{tx}}$ and $\varrho_{\mathrm{rx}}$ are trade-off factors. For $\mathbb{P}(RX\,|\,TX)$, $\mathbb{P}(TX)$ and

$\mathbb{P}(RX)$, we can use the method of deriving $\hat{\mathbb{P}}(TX_T | RX_T)$ above and use historical data for approximation. $\mathcal{U}_{\text{tx}}(TX)$ is defined as follows:

$$\mathcal{U}_{\text{tx}}(TX) = \mathcal{U}_{\text{tx}}(u,p) = \frac{p - \tau_d(u)}{2^{u-7} \tau_s(u)}. \tag{5}$$

$\mathcal{U}_{\text{tx}}(u,p)$ considers the inherent characteristics of the two key factors, the uplink spreading factor and the transmit power, to form a comprehensive indicator to provide prior knowledge of the LoRa link. The air time of a LoRa message is exponentially increasing with the spreading factor used. For every increase of 1 in the spreading factor, the air time doubles. Therefore, the exponential decay term $1/(2^{u-7})$ in $\mathcal{U}_{\text{tx}}(u,p)$ reflects the impact of the air time of the spreading factor on the message success rate. The longer the air time, the greater the probability of wireless signal collision between the message and other messages, and the lower the message success rate. The second term in $\mathcal{U}_{\text{tx}}(u,p)$ reflects the impact of the transmit power and spreading factor characteristics on the message success rate. $p - \tau_d(u)$ is the power margin term, where $p$ represents the actual transmit power, and $\tau_d(u)$ is the receiving sensitivity threshold corresponding to the spreading factor $u$. The difference between the two, $p - \tau_d(u)$, represents the "margin power" or "excess power" of the transmit power relative to the minimum sensitivity requirement. $\tau_s(u)$ is the demodulation threshold corresponding to the spreading factor $u$. The larger the value, the stronger the signal's anti-interference ability and the higher the message success rate. $\mathcal{U}_{\text{rx}}(RX)$ is defined as follows:

$$\mathcal{U}_{\text{rx}}(RX) = \mathcal{U}_{\text{rx}}(d,w) = \frac{w(1 - \tau_d(d))}{2^{d-7} \tau_s(d)}. \tag{6}$$

Similar to $\mathcal{U}_{\text{tx}}(u,p)$, $\mathcal{U}_{\text{rx}}(d,w)$ considers the inherent characteristics of the downlink spreading factor and the receiving window size to provide prior knowledge of the LoRa link. The denominator of the first term $1/(2^{d-7})$ also reflects that the longer the air time, the lower the message success rate. The numerator of the first term $w$ indicates that the larger the receiving window, the higher the probability of the downlink message being successfully received. The second term $(1 - \tau_d(d))/\tau_s(d)$ is similar to $\mathcal{U}_{\text{tx}}(u,p)$, reflecting the impact of the spreading factor characteristics on the message success rate.

Combined with the above analysis, in scenarios where data is limited or it is difficult to collect a large amount of data, the optimization problem $\mathbf{P}_2$ can introduce prior knowledge such as domain experience to overcome the problems caused by insufficient data. In addition, it can also improve the interpretability and robustness of the model. Therefore, by introducing prior knowledge into the optimization problem $\mathbf{P}_2$, the final optimization problem $\mathbf{P}_3$ is obtained:

$$\mathbf{P}_3: \underset{\{u,p\}}{\operatorname{argmax}} \; \hat{\mathbb{P}}(d,w | u,p) \frac{\hat{\mathbb{P}}(u,p) + \varrho_{\text{tx}} \cdot \mathcal{U}_{\text{tx}}(u,p)}{\hat{\mathbb{P}}(d,w) + \varrho_{\text{rx}} \cdot \mathcal{U}_{\text{rx}}(d,w)} \tag{7}$$

$$\text{s.t. (a)}: \{U = u \,|\, u \in \mathcal{A}_{USF}\},$$

$$\text{(b)}: \{P = p \,|\, p \in \mathcal{A}_P\}.$$

## 2.2 Online Cloud-Edge Knowledge Distillation

The edge decision-making method proposed in the previous section only relies on prior knowledge and historical data. In order to improve the decision-making performance of the edge decision maker in the global environment, it is also possible to consider the interaction between the edge decision maker and the cloud model, such as federated learning, segmented learning, knowledge distillation and other methods. Among them, federated learning and segmented learning have obvious advantages in protecting data privacy and distributed training, but they have high requirements for communication and energy, which makes their application in resource-constrained IoT environments face many challenges. In federated learning, each IoT device needs to regularly upload the model parameters or gradients obtained from local training to the cloud and receive global model updates at the same time. Segmented learning requires uploading intermediate activation values to the cloud and receiving gradient information sent by the cloud model. Since current deep learning models usually have many parameters, for IoT networks with limited bandwidth, this frequent, two-way non-business data exchange will significantly increase the communication load, which is easy to cause network congestion and increased latency.

The knowledge distillation method is a method that aims to transfer knowledge from a large deep neural network model (also known as a teacher model) to a lightweight network model (also known as a student model). Compared with federated learning and segmented learning, the knowledge distillation method only needs to pass the compressed teacher model output, which is usually much less than the complete model parameters, reducing the communication overhead of training. In addition, the lightweight model obtained by knowledge distillation is easier to run independently on resource-constrained IoT devices, which can not only improve the operating efficiency but also extend the battery life of the device. Therefore, in resource-constrained IoT networks, knowledge distillation of cloud models and local models of nodes can often better balance communication constraints and energy constraints.

In the simplest form of the knowledge distillation method, the student model learns by imitating the output of the teacher model on the dataset. The following loss function is often used to train the student model:

$$\mathcal{L}_{\mathrm{KL}}\left(\mathrm{S}(z_n)||\mathrm{S}(v_n)\right) = \sum_{k=1}^{K} \mathrm{S}(z_n)^{(k)} \log\left(\frac{\mathrm{S}(z_n)^{(k)}}{\mathrm{S}(v_n)^{(k)}}\right). \tag{8}$$

Where $\mathrm{S}(x) = \exp(x/\mathcal{T})/\mathrm{sum}(\exp(x/\mathcal{T}))$ is a softmax operation with a temperature coefficient $\mathcal{T}$ on the x vector. Suppose the current dataset $\mathfrak{D} = \{x_n | n = 1, \cdots, N\}$ contains $N$ samples. $K$ is the dimension of the logit vector, which is usually equal to the number of categories in classification tasks. The upper right subscript $(k)$ of the vector is the $k$th element of the vector. Given an input $x_n$, the logit vectors output by the teacher model $\mathcal{M}_\mathrm{T}$ and the student model $\mathcal{M}_\mathrm{S}$ are $v_n$ and $z_n$, respectively, that is, $v_n = \mathcal{M}_\mathrm{T}(x_n)$ and $z_n = \mathcal{M}_\mathrm{S}(x_n)$. However, Sun et al. [1] pointed out that using the same and fixed $\mathcal{T}$ for the teacher model and the student model will lead to logit shift and variance matching problems, and proposed a logit standardization method to solve this problem. This paper improves the logit standardization method and enables the cloud-based teacher model to include more information in the features sent down without increasing the model communication cost, thereby improving the efficiency of knowledge distillation.

Define the transformation functions $f_z(z_n)$ and $f_v(v_n)$ of $v_n$ and $z_n$. $F(x) = \exp(x)/\text{sum}(\exp(x))$ is the softmax function. $f_z(z_n)^{(k)} = (z_n^{(k)} + \kappa_{n,1}^{(k)})/\kappa_{n,2}^{(k)}$, $f_v(v_n)^{(k)} = (v_n^{(k)} + \kappa_{n,3}^{(k)})/\kappa_{n,4}^{(k)}$, where $\kappa_{n,1}^{(k)}$, $\kappa_{n,2}^{(k)}$, $\kappa_{n,3}^{(k)}$ and $\kappa_{n,4}^{(k)}$ are the coefficients corresponding to the $k$ th component of the logit vector of sample $n$. $v_n$ and $z_n$ are transformed by $F_z(\cdot)$ and $F_v(\cdot)$ respectively and then passed through the softmax function as follows:

$$F(f_z(z_n))^{(k)} = \frac{\exp((z_n^{(k)} + \kappa_{n,1}^{(k)})/\kappa_{n,2}^{(k)})}{\sum_{m=1}^{K} \exp((z_n^{(m)} + \kappa_{n,1}^{(m)})/\kappa_{n,2}^{(m)})}. \tag{9}$$

$$F(f_v(v_n))^{(k)} = \frac{\exp((v_n^{(k)} + \kappa_{n,3}^{(k)})/\kappa_{n,4}^{(k)})}{\sum_{m=1}^{K} \exp((v_n^{(m)} + \kappa_{n,3}^{(m)})/\kappa_{n,4}^{(m)})}. \tag{10}$$

For an ideal student model, its KL divergence loss is 0, that is, $\forall k \in [1,K]$, satisfying $F(f_z(z_n))^{(k)} = F(f_v(v_n))^{(k)}$. Then we can get:

$$\frac{\exp(f_z(z_n)^{(k)})}{\sum_{m=1}^{K} \exp(f_z(z_n)^{(m)})} = \frac{\exp(f_v(v_n)^{(k)})}{\sum_{m=1}^{K} \exp(f_v(v_n)^{(m)})}$$

$$\exp(f_z(z_n)^{(k)}) = \frac{\sum_{m=1}^{K} \exp(f_z(z_n)^{(m)})}{\sum_{m=1}^{K} \exp(f_v(v_n)^{(m)})} \exp(f_v(v_n)^{(k)})$$

$$f_z(z_n)^{(k)} = \log\left(\frac{\sum_{m=1}^{K} \exp(f_z(z_n)^{(m)})}{\sum_{m=1}^{K} \exp(f_v(v_n)^{(m)})}\right) + f_v(v_n)^{(k)} \tag{11}$$

$$\frac{z_n^{(k)} + \kappa_{n,1}^{(k)}}{\kappa_{n,2}^{(k)}} = \log\left(\frac{\sum_{m=1}^{K} \exp(f_z(z_n)^{(m)})}{\sum_{m=1}^{K} \exp(f_v(v_n)^{(m)})}\right) + f_v(v_n)^{(k)}.$$

Taking the derivative of $z_n^{(k)}$ with respect to $\kappa_{n,3}^{(k)}$ yields:

$$\frac{\partial z_n^{(k)}}{\partial \kappa_{n,3}^{(k)}} = \frac{\kappa_{n,2}^{(k)}}{\kappa_{n,4}^{(k)}} \left(1 - F(f_v(v_n))^{(k)}\right). \tag{12}$$

Since $1 - F(f_v(v_n))^{(k)} > 0$, the update direction and step size of $z_n^{(k)}$ can be guided by adjusting $\kappa_{n,2}^{(k)}$, $\kappa_{n,3}^{(k)}$, and $\kappa_{n,4}^{(k)}$. If the coefficients of the logit elements of the same sample are all equal, this method degenerates into the logit standardization method, that is, $\forall k \in [1,K]$, let $\kappa_{n,1}^{(k)} = -\overline{z_n}$, $\kappa_{n,2}^{(k)} = \sigma(z_n)$, $\kappa_{n,3}^{(k)} = -\overline{v_n}$, $\kappa_{n,4}^{(k)} = \sigma(v_n)$. Substituting it into the derivative yields:

$$\left.\frac{\partial z_n^{(k)}}{\partial \kappa_{n,3}^{(k)}}\right|_{\kappa_{n,3}^{(k)} = -\overline{v_n}} = \frac{\sigma(z_n)}{\sigma(v_n)}\left(1 - \frac{\exp(v_n^{(k)}/\sigma(v_n))}{\sum_{m=1}^{K} \exp(v_n^{(m)}/\sigma(v_n))}\right). \tag{13}$$

It can be seen that if the coefficient $\kappa_{n,3}^{(k)}$ of the logit element of the same sample is fixed to any constant, that is, $\kappa_{n,3}^{(k)} = \kappa_{n,3}$, the gradient information obtained is the same. Therefore, the logit elements of the same sample can have different $\kappa_{n,3}^{(k)}$, which can be used to provide more information for the student model to speed up the knowledge distillation from teacher to student.

Combined with the LoRaWAN network, the distillation process needs to be carried out after the terminal node receives the control message sent by the gateway. However, due to network conflicts and the sparseness of node traffic, the time interval for the terminal node to receive the control message sent by the gateway may be long and the model deployed in the cloud is continuously updated, which causes the update of the edge model to lag. To this end, the experience information of other nodes can be introduced into $\kappa_{n,3}^{(k)}$ to make the edge model more robust. Let $\kappa_{n,3}^{(k)} = \mathbb{E}_{e \sim \mathcal{D}}[h(e, \mathrm{x}_n^{(k)})]$, where $\mathcal{D}$ is the set of all nodes in the LoRaWAN network, $h(e, \mathrm{x}_n^{(k)})$ is the historical cumulative success rate of the terminal node $e$ sending messages in the state $\mathrm{x}_n^{(k)}$. Similar to logit normalization, $\forall k \in [1, K]$, let $\kappa_{n,1}^{(k)} = -\overline{\mathrm{z}_n}$, $\kappa_{n,2}^{(k)} = \sigma(\mathrm{z}_n)$, $\kappa_{n,4}^{(k)} = \sigma(\mathrm{v}_n)$, substitute $f_\mathrm{z}(\mathrm{z}_n)$ and $f_\mathrm{v}(\mathrm{v}_n)^{(k)}$ can be obtained:

$$\begin{aligned} f_\mathrm{z}(\mathrm{z}_n)^{(k)} &= \left(\mathrm{z}_n^{(k)} - \overline{\mathrm{z}_n}\right)\big/\sigma(\mathrm{z}_n), \\ f_\mathrm{v}(\mathrm{v}_n)^{(k)} &= \left(\mathrm{v}_n^{(k)} + \mathbb{E}_{e \sim \mathcal{D}}[h(e, \mathrm{x}_n^{(k)})]\right)\big/\sigma(\mathrm{v}_n). \end{aligned} \quad (14)$$

This method introduces the information of all other nodes in the LoRaWAN network into the training process. When the student model approaches the current teacher model, the optimization direction of the student model also considers the historical cumulative success rate of other nodes in this state. Therefore, the efficiency of knowledge distillation can be improved without increasing any communication overhead.

## 2.3 Intelligent Downlink Control and Load Balancing Based on Lyapunov Optimization

In large-scale LoRaWAN networks, multiple gateways are often used to improve the success rate of uplink messages and load balance the downlink messages to avoid overloading the downlink channel of a single gateway. Although using multiple LoRaWAN gateways can enhance the downlink capacity of the network, a LoRaWAN gateway has 8 uplink channels and only 1 downlink channel. As the scale of the network expands, the number of uplink messages will gradually exceed the upper limit of the downlink capacity of the LoRaWAN gateway, which will inevitably lead to a large number of uplink messages being unable to receive downlink control messages from the gateway. Therefore, in large-scale multi-gateway LoRaWAN networks, it is crucial to formulate reasonable intelligent downlink control and load balancing strategies for massive uplink messages. For the intelligent downlink control and load balancing strategy, the specific problems to be solved can be described as: Under the condition of limited downlink capacity of the gateway, how to select appropriate uplink messages to reply to downlink control messages to maximize network performance; and how to assign corresponding gateways to these downlink control messages to achieve load balancing? Next, these problems are analyzed one by one and solutions are proposed.

Here, we ignore the condition that the gateway's downstream capacity is limited and analyze the decision-making problem. For the intelligent downstream control and load balancing strategy, the decision process can be modeled as a Markov decision process and solved using the Actor-Critic algorithm architecture. Define the gateway set $\mathcal{G} = \{g_i | i = 1, \cdots, G\}$, where $G$ is the number of gateways. The downstream control message queue of gateway $g$ in time slot $t$ is $\mathcal{B}_g(t) = \{b_i | i = 1, \cdots, B\}$, where $B$ is the number of downstream control messages. For a downlink control message $b$, its state in the Markov decision process is expressed as $s_b = s_b^* \cup a_b^* \cup \{l_{g_i} | i = 1, \cdots, G\} \cup \{\text{TOA}(b)\}$. Among them, $s_b^*$ and $a_b^*$ are the state of the target terminal node of the downlink control message $b$ and the control instruction issued by the gateway. $l_{g_i}$ is the current downlink control message queue length of the gateway $g_i$. $\text{TOA}(b)$ is the air time of the downlink control message $b$. The action space of this decision problem is $a_b = \{\mathcal{W}\}$, $\mathcal{W}$ represents the gateway number that sends the downlink control message. If $\mathcal{W} = G$, the downlink decision message is not sent for the current node. The reward function $r_b$ of this decision problem is defined as follows:

$$r_b(s_b, a_b) = \tan\left(\text{sgn}(\mathcal{W} < G) \cdot \Delta \text{H}(s_b, a_b)\right) + \mathbb{I}(\mathcal{W} < G) \cdot l_{g_i}. \tag{15}$$

Among them, $\Delta \text{H}(s_b) = \text{H}(a_b^*) - \text{H}(s_b^*)$, where $\text{H}(a_b^*)$ is the global cumulative PDR after the terminal node executes the action $a_b^*$, and $\text{H}(s_b^*)$ is the global cumulative PDR when the terminal node maintains the state $s_b^*$. $\text{H}(\cdot)$ is the global cumulative PDR under the given parameters. $\Delta \text{H}(s_b)$ reflects the global cumulative PDR change before and after the terminal node executes the action $a_b^*$, which is used to evaluate the long-term return of the action $a_b^*$, thereby reducing short-term network fluctuations. The trade-off factor $\lambda$ is used to balance the instantaneous PDR and the global cumulative PDR change. The application of negative logarithmic transformation increases the penalty for actions with low success rate, aiming to improve the training efficiency of the entire network.

In terms of network structure, the state-action pair value function $Q_{\varphi_l}^\pi(s, a)$ and the policy function $\pi_{\theta_l}(a|s)$ are defined, where $\varphi_l$ and $\theta_l$ are the parameters of their respective functions. Optimize $Q_{\varphi_l}^\pi(s, a)$ by minimizing the mean square error between $Q_{\varphi_l}^\pi(s, a)$ and its temporal difference target. For the policy function $\pi_{\theta_l}(a|s)$, maximize the policy distribution weighted by the state-action pair value. The specific objective functions are:

$$\min_{\varphi_l} \mathbb{E}_{S_l \sim \mathcal{B}_l}\left[\left(r_b + \gamma \max_q Q_{\varphi_l}^\pi(s_b', q) - Q_{\varphi_l}^\pi(s_b, a_b)\right)^2\right], \tag{16}$$

$$\max_{\theta_l} \mathbb{E}_{s_b \sim \mathcal{B}_l}\left[\mathbb{E}_{a_b \sim \pi_{\theta_l}}[Q_{\varphi_l}^\pi(s_b, a_b) \cdot \log \pi_{\theta_l}(a_b | s_b)]\right]. \tag{17}$$

Where $\mathcal{B}_l$ is the experience replay cache, and the new samples $S_l = (s_b, a_b, s_b', r_b)$ collected during the interaction with the environment are included in $\mathcal{B}_l$.

The Actor-Critic based reinforcement learning method proposed above can already handle decision-making problems, but it does not take into account constraints. Next, we will analyze the conditions under which the gateway's downlink capacity is limited and introduce this constraint into the proposed reinforcement learning method. $L_g(t, \pi_{\theta_l}) = \sum_{i=1}^B \text{TOA}(b_i)$ is the load added by gateway $g$ in time slot $t$ according to strategy $\pi_{\theta_l}$, where $\text{TOA}(b_i)$ is the air time of message $b_i$. The long-term average load of gateway $g$ is:

$$\overline{L_g} = \lim_{T \to \infty} \frac{1}{T} \sum_{t=0}^{T-1} \mathbb{E}[L_g(t, \pi_{\theta_l})]. \tag{18}$$

The long-term average load of each gateway needs to satisfy $\overline{L_g} \leqslant T_l$, where $T_l$ is the size of a time slot. The optimization objective function of gateway $g$ in time slot $t$ is $V_g(t, \pi_{\theta_l}) = \sum_{b=1}^{B} Q_{\varphi_l}^{\pi}(s_b, \pi_{\theta_l}(s_b))$, which reflects the sum of the state-action pair values of all downlink control messages sent by gateway $g$ in time slot $t$. The long-term average decision value of all gateways in the network is:

$$\overline{V} = \lim_{T \to \infty} \frac{1}{T} \sum_{t=0}^{T-1} \sum_{g=1}^{G} \mathbb{E}[V_g(t, \pi_{\theta_l})]. \tag{19}$$

Therefore, the problem can be further described as solving a strategy that can maximize the long-term average decision value $\overline{V}$ under the constraint condition $\overline{L_g} \leqslant T_l$ of the long-term average load of each gateway. The problem can be expressed as:

$$\mathbf{P}_4: \min_{\pi_{\theta_l}} \lim_{T \to \infty} \frac{1}{T} \sum_{t=0}^{T-1} \sum_{g=1}^{G} -\mathbb{E}[V_g(t, \pi_{\theta_l})] \tag{20}$$

$$\text{s.t.} \quad \lim_{T \to \infty} \frac{1}{T} \sum_{t=0}^{T-1} \mathbb{E}[L_g(t, \pi_{\theta_l})] \leqslant T_l, \quad g \in \{1, ..., G\}.$$

Problem $\mathbf{P}_4$ is a long-term stochastic optimization problem. In addition, solving the optimal strategy requires complete future information, which is impossible. To overcome this problem, this paper adopts the Lyapunov optimization method [2]. This method can solve a stochastic optimization problem measured from a long-term perspective in each time slice and ensure the satisfaction of long-term constraints through the stability of the virtual queue. In addition, the algorithm does not require system status information from the future to define a downlink overload queue $\mathcal{O}_g(t)$ with an initial value of 0 for each gateway:

$$\mathcal{O}_g(t+1) = \max\{\mathcal{O}_g(t) + L_g(t, \pi_{\theta_l}) - T_l, 0\}, \quad g \in \{1, ..., G\}. \tag{21}$$

In time slot $t+1$, the total load of gateway $g$ is equal to the unfinished load of the previous time slot plus the newly introduced load of the current time slot, that is, $\mathcal{O}_g(t) + L_g(t, \pi_{\theta_l})$. In time slot $t+1$, $\mathcal{O}_g(t) + L_g(t, \pi_{\theta_l}) - T_l$ is the total load minus the load that can be completed in the current time slot, indicating the unfinished overload. $\max\{\cdot, 0\}$ indicates that the load processing capacity in each time slot cannot be transferred to the next time slot, and only the unfinished overload can be transferred to the next time slot. Define the downlink overload queue set in the $t$ time slot as $\mathbf{\Theta}(t) = \{\mathcal{O}_1(t), ..., \mathcal{O}_G(t)\}$. The Lyapunov function and Lyapunov drift are:

$$L(\mathbf{\Theta}(t)) = \frac{1}{2} \sum_{g=1}^{G} \mathcal{O}_g(t)^2. \tag{22}$$

$$\Delta L(\mathbf{\Theta}(t)) = L(\mathbf{\Theta}(t+1)) - L(\mathbf{\Theta}(t)). \tag{23}$$

Among them, $L(\mathbf{\Theta}(t))$ reflects the sum of the downlink overload of all gateways in the network in time slot $t$. According to the Lyapunov optimization theory, this paper uses the Lyapunov Drift-Plus-Penalty Algorithm [2] to solve the problem

**P**$_4$. The optimal control strategy for time slot $t$ is solved as follows:

$$\min_{\theta_l} \mathbb{E}\left[\sum_{g=1}^{G} \mathcal{O}_g(t)(L_g(t,\pi_{\theta_l}) - T_l) - \lambda_l V_g(t,\pi_{\theta_l}) \middle| \Theta(t)\right]. \tag{24}$$

Among them, $\lambda_l$ is a trade-off factor. A larger $\lambda_l$ makes the solution process pay more attention to the maximization of the long-term average decision value $\overline{V}$, and a smaller $\lambda_l$ makes the solution process comply with the constraint $\overline{L_g} \lesssim T_l$ faster. In order to introduce this constraint into the proposed reinforcement learning strategy optimization process, here we set $\lambda_l = 0$, then the solution problem is transformed into:

$$\max_{\theta_l} \mathbb{E}\left[\sum_{g=1}^{G} \mathcal{O}_g(t)(T_l - L_g(t,\pi_{\theta_l})) \middle| \Theta(t)\right]. \tag{25}$$

Let $\mathrm{R}(\pi_{\theta_l}) = \sum_{g=1}^{G} \mathcal{O}_g(t)(T_l - L_g(t,\pi_{\theta_l}))$, and substitute it into the reinforcement learning strategy optimization process to obtain the strategy solution objective function considering the constraints, as follows:

$$\max_{\theta_l} \mathbb{E}_{s_b \sim \mathcal{B}_l}\left[\mathbb{E}_{a_b \sim \pi_{\theta_l}}\left[(Q_{\varphi_l}^{\pi}(s_b, a_b) + \mathrm{R}(\pi_{\theta_l}))\log \pi_{\theta_l}(a_b | s_b)\right]\right]. \tag{26}$$

# Experimental Evaluation and Results Analysis

This section uses a series of experiments to demonstrate the effectiveness of the edge intelligence-based cloud-edge collaborative resource allocation and decision-making method proposed in this chapter. It mainly includes: a horizontal comparison with existing learning methods and ablation verification of each module of the proposed method. Compared with Chapter 4, the number of LoRaWAN gateways in the following experiments is increased to 3, and the node scale is expanded to 1300 to verify the effectiveness of the proposed method in a multi-gateway LoRaWAN network.

## 3.1 Experimental Setting

In order to demonstrate the advantages of the HEAT-LDL algorithm over other methods, this chapter selects several representative methods and includes them in the comparison model for comparison under different node density and traffic conditions. They mainly include the ADRx [3], A2C [4], Qmix [5] and QUCBMAB [6] algorithms. In addition, this chapter also includes the HEAT algorithm proposed in Chapter 4 in the comparison scope.

(1) Comparative experimental results under different node densities

This part of the work evaluates the performance of the proposed HEAT-LDL algorithm and its comparative algorithms at different node densities. Fig.2 shows the changes in PDR and EER when the number of nodes $N$ increases from 800 to 1300 in a LoRaWAN network with a traffic intensity $\delta$ of 2.

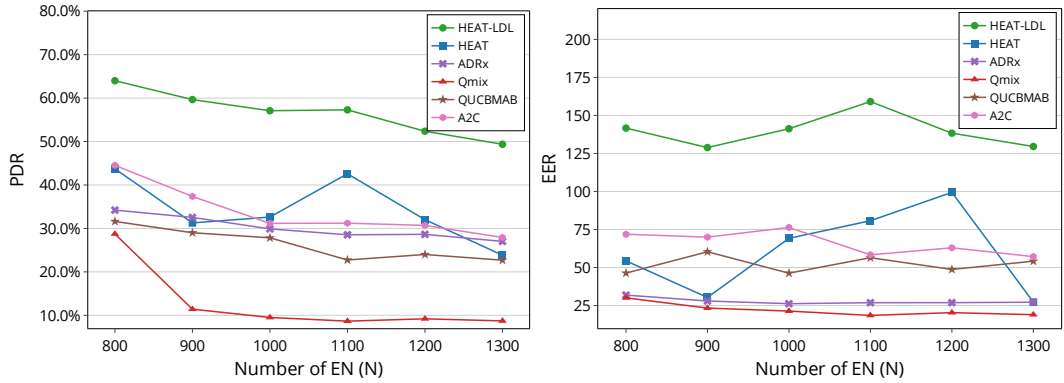

Fig.2 PDR and EER at different node densities

HEAT-LDL achieved the highest PDR at all node densities, decreasing from 64% at 800 nodes to 49.4% at 1,300 nodes. Specifically, HEAT-LDL's PDR increased by an average of 29.2% in each scenario, and an average of 20.5% higher than the best results of other algorithms. This result shows that HEAT-LDL has good robustness and adaptability in multi-gateway large-scale LoRaWAN networks, and can still ensure successful data packet transmission even when the number of nodes increases, network competition and interference increase. The PDRs of HEAT and A2C start at 43.7% and 44.5%, respectively, which are lower than HEAT-LDL overall. Both of them show certain fluctuations at different node densities, indicating that their strategies may have certain instability or lack of adaptability under specific network loads. In addition, ADRx, Qmix and QUCBMAB are at a relatively low level overall and change relatively smoothly with the change of node density. Overall, these methods are relatively lacking in adaptability to rapid changes in network status when dealing with resource allocation problems in large-scale networks.

The HEAT-LDL algorithm shows a relatively gentle downward trend, indicating that its design can still maintain good performance when the node density increases. This is mainly due to its edge autonomous decision-making characteristics and the load balancing module in the cloud, which enables the network to still share resource pressure well under high load conditions. HEAT and A2C: Although the starting values are close, HEAT fluctuates when the node density ranges from 900 to 1100, while A2C shows a more uniform downward trend, indicating that both have critical points of strategic advantages or disadvantages at certain node densities. Qmix is at a low level at the initial node density and quickly reaches a plateau as the node density increases, indicating that it can hardly effectively improve PDR in extremely large-scale networks.

From the experimental results, the EER of HEAT-LDL at different node densities is significantly higher than that of other algorithms, achieving better network resource utilization efficiency. Specifically, the EER of HEAT-LDL increased by an average of 285.1% in each scenario, which is an average increase of 88.1% over the best results of other algorithms. This shows that HEAT-LDL can more effectively allocate resources and control power in a large-scale multi-gateway LoRaWAN environment, thereby improving transmission efficiency. Secondly, the EER of HEAT and A2C is acceptable at some node densities, but the overall fluctuation is large, especially the energy efficiency ratio of HEAT at different node densities fluctuates greatly, showing a certain instability. The EER of ADRx, Qmix and QUCBMAB is low, especially Qmix, the energy efficiency ratio continues to decline when the number of nodes increases, indicating that it has poor adaptability in a large-scale LoRaWAN environment.

(2) Comparative experimental results under different traffic intensities

This part of the work evaluates the performance of the proposed HEAT-LDL

algorithm and its comparative algorithm in scenarios with different traffic intensities. Fig.3 shows the changes in PDR and EER when the traffic intensity $\delta$ increases from 0.5 to 3 in a LoRaWAN network with 1000 terminal nodes $N$.

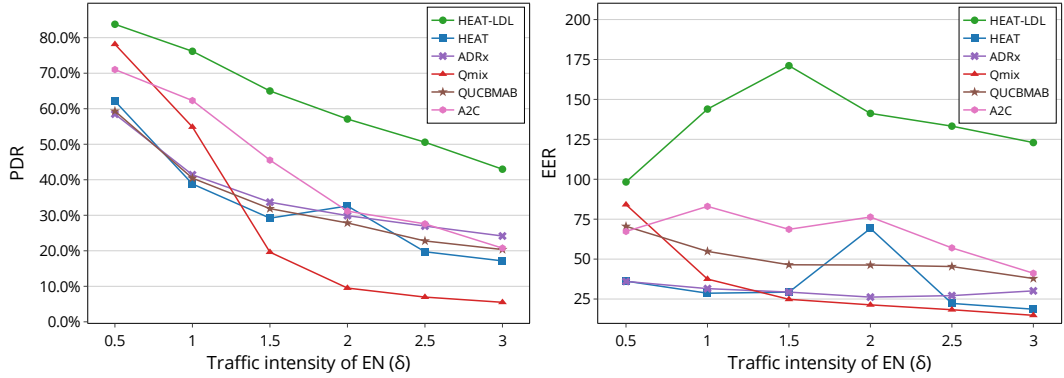

Fig.3 PDR and EER under different traffic intensities

As the traffic intensity increases from 0.5 to 3, the PDR of each algorithm shows a downward trend, which to some extent reflects the challenges of increased interference, conflict and resource competition in the network under high traffic conditions. The PDR of HEAT-LDL increased by an average of 27.6% in each scenario, which is an average of 17.6% higher than the best results of other algorithms. When the traffic intensity is low, the PDR of the HEAT-LDL algorithm can reach 75% to 85%, which is significantly higher than other comparison algorithms. This shows that its edge decision module based on prior knowledge and historical data can help nodes make better choices when the information is incomplete. As the traffic intensity increases, the PDR of HEAT-LDL decreases slowly, but it still remains at the optimal level, while other methods all decrease to a lower level.

Qmix regards each terminal as an independent agent and realizes collaborative decision-making through multi-agent reinforcement learning. It is competitive with HEAT-LDL at low traffic intensity, but it drops sharply at medium and high traffic intensity. This is because the multi-agent coordination problem is more prominent under high load, resulting in low overall decision efficiency. A2C performs well when the traffic intensity is 0.5 to 1.5, with a PDR of 71% to 62.3%, but the PDR also decreases significantly as the traffic increases. This means that although the Actor-Critic framework has good stability in continuous decision-making problems, pure online decision-making lacks additional edge and cloud collaborative information, and may not be able to adjust strategies in time to cope with dynamic changes in high-load scenarios.

From the experimental data, different algorithms show different trends and fluctuations when the traffic intensity changes from 0.5 to 3. Among them, the HEAT-LDL algorithm has the highest overall performance among all algorithms, and reaches a peak when the traffic is 1.5; while other algorithms show varying degrees of attenuation or fluctuation. Specifically, the EER of HEAT-LDL increased by an average of 298.7% in each scenario, which is an average of 109.5% higher than the best results of other algorithms. HEAT-LDL gradually increased from 98.297 at low traffic to 171.114 at 1.5 traffic, and then decreased to 122.963 after the traffic increased further. Under low traffic intensity conditions, the terminal nodes send data less frequently, and the model learns the optimal strategy through data more slowly, resulting in a lower EER. HEAT-LDL can fully utilize resources under low to medium traffic conditions. When the traffic is at a moderate level, the network congestion is not serious, and its scheduling or routing strategy can achieve higher energy efficiency; when the traffic is too large, network competition intensifies, and the energy efficiency ratio begins to decline.

## 3.2 Ablation Experiment

In order to explore the role of the different modules proposed in the previous section, this section conducts ablation verification and analysis on the proposed HEAT-LDL algorithm. Among them, "only local" means that only the edge decision module based on prior knowledge and local history is used; "only distill" means that only the online cloud-edge knowledge distillation module is used; "only Ly" means that only the intelligent downlink control and load balancing module based on Lyapunov optimization is used; "local+distill" means that both the edge decision module based on prior knowledge and local history and the online cloud-edge knowledge distillation module are used.

(1) Ablation experiment results under different node densities

This part of the work evaluates the performance of the proposed HEAT-LDL algorithm and its ablation algorithm at different node densities. Fig.4 shows the changes in PDR and EER when the number of nodes $N$ increases from 800 to 1300 in a LoRaWAN network with a traffic intensity $\delta$ of 2.

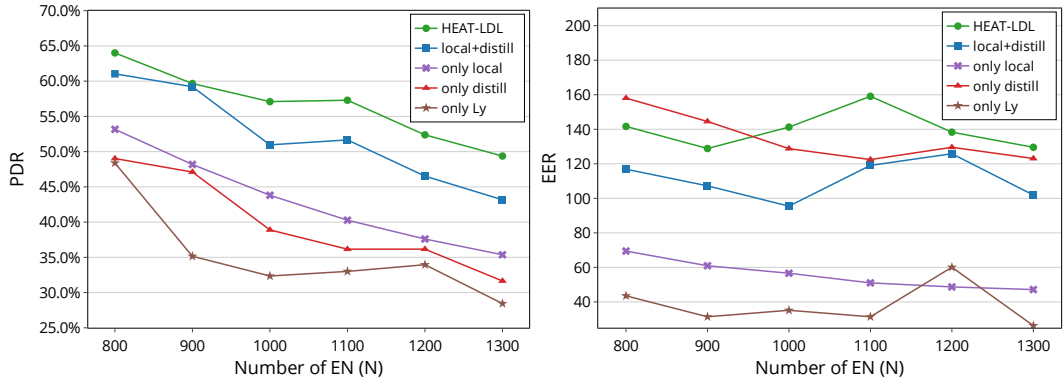

Fig.4 PDR and EER at different node densities

As the number of nodes increases from 800 to 1300, HEAT-LDL performs best at all node densities. Specifically, HEAT-LDL improves PDR by an average of 14% compared to other ablation schemes, and improves by an average of 4.33% relative to the best result of the ablation scheme. Although the overall phenomenon of the "local+distill" combination is similar to that of HEAT-LDL, its PDR is slightly lower, especially when the node density is high, the performance degradation is more obvious. Among the individual module schemes, "only local", "only distill" and "only Ly" all have low PDR, among which "only local" is between "only distill" and "only Ly", and "only Ly" is particularly unsatisfactory at high density, with only 33% at 1100 nodes and 28.4% at 1300 nodes.

In terms of EER, HEAT-LDL presents a relatively stable energy efficiency level, with values fluctuating between 129 and 159, indicating that the transmission efficiency of terminal nodes can be well guaranteed at different node densities. The EER of HEAT-LDL increased by an average of 120.6% in each scenario, which is 5.1% higher than the best result of the ablation scheme. "Only distill" showed a higher EER of 158.05 and 144.52 respectively when the node density was low, but its advantage gradually weakened as the node density increased, and the gap with HEAT-LDL narrowed or even slightly lower. The EER values of "only local" and "only Ly" were generally low, and there was a more obvious energy efficiency drop when the node density increased, indicating that modules that rely solely on edge history or Lyapunov control are difficult to balance high efficiency and transmission success rate.

(2) Ablation experiment results at different flow intensities

This part of the work evaluates the performance of the proposed HEAT-LDL algorithm and its ablation algorithm in scenarios with different traffic intensities. Fig.5 shows the changes in PDR and EER when the traffic intensity $\delta$ increases from 0.5 to 3 in a LoRaWAN network with 1000 terminal nodes $N$.

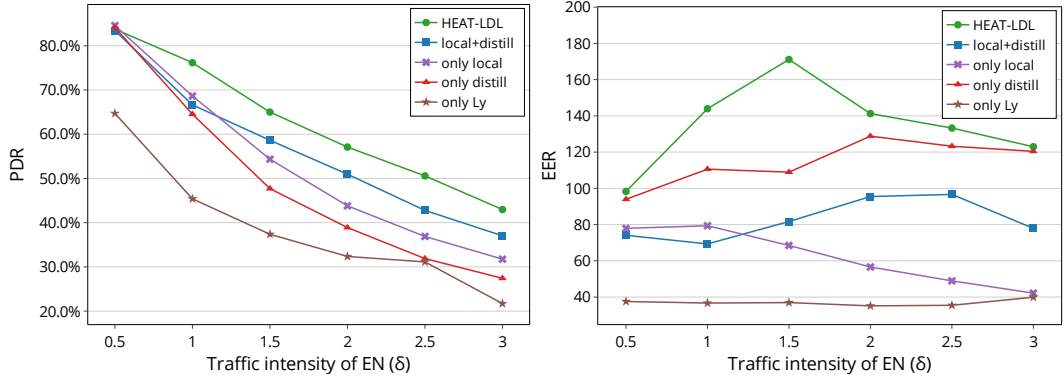

Fig.5 PDR and EER under different traffic intensities

When the traffic intensity is 0.5, the PDR of HEAT-LDL is comparable to other ablation schemes, and slightly lower than 84.6% of "only local" and 84.1% of "only distill". However, as the traffic intensity increases, other ablation schemes have a significant performance decline, while HEAT-LDL's decline is gentle and at the highest level. The PDR of HEAT-LDL has increased by an average of 13.2% in each scenario, and an average of 5.5% higher than the best results of other algorithms. When the traffic reaches 3, the PDR of HEAT-LDL remains at 43%, which is significantly higher than other schemes. This shows that in a high-traffic environment, HEAT-LDL, which integrates three modules, is more resistant to interference and resource competition and maintains a high data transmission success rate.

HEAT-LDL shows a trend of first rising and then falling in EER: from 98.297 at 0.5 to 171.114 at 1.5, and then gradually decreases as the traffic increases further. Overall, HEAT-LDL can significantly improve the transmission efficiency per unit energy under medium-intensity traffic, and still maintains its advantages under low and high traffic intensities. The EER of HEAT-LDL increased by an average of 120% in each scenario, and was 18.6% higher than the best result of the ablation scheme on average. The performance of "only distill" is similar to that of HEAT-LDL when the traffic intensity is low and high, but there is a large gap with HEAT-LDL when the traffic intensity is medium. The EER performance of "local+distill" is always lower than that of HEAT-LDL, indicating that the energy utilization efficiency is difficult to further improve after the global load balancing regulation of the cloud is lost. The EER of "only local" and "only Ly" is low regardless of the traffic, especially "only Ly" has the weakest energy efficiency, which proves that a single strategy is difficult to optimize both the transmission success rate and energy consumption.

# Conclusion

In this chapter, a cloud-edge collaborative resource allocation and decision-making method based on edge intelligence, HEAT-LDL, is proposed to solve the downlink starvation problem caused by asymmetric uplink and downlink resources in large-scale multi-gateway LoRaWAN networks. By performing online knowledge distillation on the cloud model at the edge and using the edge decision-making method that integrates prior knowledge and local historical data at the edge, autonomous collaborative decision-making at the edge is achieved, effectively solving key problems such as decision information lag caused by insufficient downlink capacity of the gateway. At the same time, for the downlink message control problem between multiple gateways, HEAT-LDL combines the Lyapunov optimization method with the Actor-Critic decision-making method, and ensures the satisfaction of long-term constraints through the virtual queue mechanism, thereby realizing intelligent downlink control and load balancing.

The experimental part verifies the superiority of HEAT-LDL under different node densities and traffic intensities through comparative experiments and ablation experiments. The results show that HEAT-LDL is significantly better than the existing algorithms in improving the overall PDR and EER of the network, and the synergy of each module further enhances the robustness of the decision and the efficiency of network resource utilization. The method proposed in this chapter not only takes into account the synergistic effect of global optimization and local decision-making, but also takes into account the scalability and energy consumption control of the system, providing a practical solution for resource allocation and decision-making in large-scale LoRaWAN networks.